\documentclass{emulateapj}
\usepackage{amsmath,amssymb,amsfonts}
\slugcomment{Accepted by The Astrophysical Journal}
\shortauthors{USCANGA ET AL.}

\begin{document}
\title{Statistical Analysis of Water Masers in Star-Forming Regions: Cepheus~A and W75~N} 

\author{L. Uscanga\altaffilmark{1}, J. Cant\'o\altaffilmark{2},  J. F. G\'omez\altaffilmark{1}, G. Anglada\altaffilmark{1}, J. M. Torrelles\altaffilmark{3},
N. A. Patel\altaffilmark{4}, A. C. Raga\altaffilmark{5}, and S. Curiel\altaffilmark{2}}

\altaffiltext{1}{Instituto de Astrof\'{\i}sica de Andaluc\'{\i}a (CSIC), Apartado 3004, E-18080 Granada, Spain; lucero@iaa.es, jfg@iaa.es, guillem@iaa.es}

\altaffiltext{2}{Instituto de Astronom\'{\i}a, Universidad Nacional Aut\'onoma 
de M\'exico, Apartado 70-264, 04510 M\'exico, D. F., Mexico; scuriel@astroscu.unam.mx}

\altaffiltext{3}{Instituto de Ciencias del Espacio (CSIC)-UB-IEEC, Facultat de F\'isica, Universitat de Barcelona, Planta 7a, Mart\'{\i} i Franqu\`{e}s 1, E-08028 Barcelona, Spain; torrelles@ieec.fcr.es}

\altaffiltext{4}{Harvard-Smithsonian Center for Astrophysics, 60 Garden Street, Cambridge, MA 02138, USA; npatel@cfa.harvard.edu}

\altaffiltext{5}{Instituto de Ciencias Nucleares, Universidad Nacional Aut\'onoma de M\'exico, Apartado 70-543, 04510 M\'exico, D. F., Mexico; raga@nucleares.unam.mx}

\begin{abstract}
We have done a statistical analysis of Very Long Baseline Array (VLBA) data of water masers in the star-forming regions (SFRs) Cepheus A and W75 N, using correlation functions to study the spatial clustering and Doppler-velocity distribution of these masers. Two-point spatial correlation functions show a characteristic scale size for clusters of water maser spots $\la$1~AU, similar to the values found in other SFRs. This suggests that the scale for water maser excitation tends to be $\la$1 AU. Velocity correlation functions show power-law dependences with indices that can be explained by regular velocity fields, such as expansion and/or rotation. These velocity fields are similar to those indicated by the water maser proper-motion measurements; therefore, the velocity correlation functions appear to reveal the organized motion of water maser spots on scales larger than 1 AU.
\end{abstract}


\section{INTRODUCTION}
Water maser emission is commonly found in star-forming regions (SFRs) within clumps of molecular gas with high densities ($\sim10^8-10^9$~cm$^{-3}$) and warm temperatures ($\sim$400~K) \citep{Eli89}. These physical conditions can be reached in both the inner parts of accretion disks around young stellar objects (YSOs) and the gas compressed by shock waves generated by outflows associated with these objects.

Water masers are very compact and intense ($\sim10^{13}$~cm, $T_b\sim10^{12}$~K), allowing studies through Very Long Baseline Interferometry (VLBI) observations to analyze in detail the spatio-kinematical distribution of the gas around YSOs with very high angular and spectral resolutions. In particular, sources with a large number of masers allow us to obtain through a statistical analysis the clustering and velocity distribution of masers from scales of a few hundred of AUs down to less than 1~AU. Statistical studies give important information about the typical size for clusters of masers that may be related to a fundamental scale of its excitation, as well as the statistical properties of both the clustering and the velocity field.

In this context, \citet{Gwi92} carried out VLBI water maser observations of the SFR W49. They identified 271 maser features with 625 maser spots, which were used for a statistical study. A maser spot is an individual component of the maser emission occurring at a given velocity channel and position. Maser spots tend to cluster in position and Doppler velocity, typically 1~AU and 0.5~km~s$^{-1}$ \citep{Gwi94a}. Such clusters or groups of maser spots are called maser features. Physically, maser features should be small clouds supporting population inversion by a pumping mechanism \citep{Gwi94b}. The statistical analysis of spatial and velocity distributions for both spots and features reveals that turbulent motions dominate on a spatial scale of $\sim$1--300~AU \citep{Gwi94a}.

\citet{Ima02} carried out a similar study using Very Long Baseline Array (VLBA) maser data of the SFR W3 IRS 5. For the statistics, \citet{Ima02} used 905 maser spots grouped in 152 maser features. They found that spots form features with a typical size of $\sim$0.5~AU. The statistical analysis of the Doppler velocities, specifically, the velocity correlation functions for maser spots follow a power-law dependence in the range of 0.04$-$300~AU with an index of $\sim$0.29, that is consistent with the Kolmogorov value of 1/3, the value expected for incompressible fluids with a turbulent velocity field \citep{Kol41,Str07}. Similar results have been also obtained by \citet{Str02} in other five SFRs.

VLBA water maser observations toward the SFRs Cepheus A and W75 N revealed remarkable microstructures \citep{Tor01b,Tor03}. These microstructures exhibit a coherent and well-ordered spatio-kinematical behavior at AU scales \citep{Usc05}. Proper-motion measurements of water masers suggest the presence of  organized motions of structures with sizes from tens to a few hundreds of AUs \citep{Tor01b,Tor03}. Here we study the spatial and velocity distribution of the water masers in these two SFRs using a statistical analysis to investigate whether organized or turbulent motions dominate over spatial scales from a few hundred of AUs down to less than 1~AU.

This paper is organized as follows. In Section 2, we define the correlation functions used in this statistical study. In Sections 3 and 4, we give a brief introduction to the SFRs, Cepheus A and W75~N, and present the results of the correlation functions, respectively. In Section 5, we discuss the clustering of water maser spots and give a possible interpretation for the velocity field traced by the water masers, based on the statistical analysis. Finally, we summarize our conclusions in Section 6.
 
\section{Statistical analysis of maser spots}
\subsection{Two-point spatial correlation function}
We adopt the definition previously used by \citet{Wal84} and \citet{Ima02}, considering the two-point spatial correlation function of maser spots as the number of spots per unit angular area with a given separation on the plane of sky, $\Delta r$, from an arbitrary spot. This function can be expressed as 
\begin{equation}
n_s(\Delta r)d\Omega=\frac{\sum_{i,j}n_{\delta}(\vert \boldsymbol{r}_i-\boldsymbol{r}_j\vert)}{n_{\mathrm{spot}}}\,,
\end{equation}
where $n_{\mathrm{spot}}$ is the total number of maser spots, the vectors \mbox{\boldmath{$r$}}$_i$ and \mbox{\boldmath{$r$}}$_j$ determine the positions of the spots, and the indices $i$, $j$ run over all spots. Here
\begin{equation}
n_{\delta}(r)=\left\{\begin{array}{ll}
1 & \textrm{when}\quad \Delta r<r<\Delta r+dr\,,\\
0 & \textrm{otherwise\,,}
\end{array}\right.
\end{equation}
where $\Delta r=(\Delta x^2+\Delta y^2)^{1/2}$, $\Delta x$ and $\Delta y$ are the offsets in the right ascension and declination axis directions, respectively; $dr$ is a value for the separation of the successive bins in the $n_s$ vs. $\Delta r$ plot, and $d\Omega=2\pi\Delta rdr+\pi(dr)^2$ is the total area of the annulus where the counting is done.

Frequently, the two-point correlation functions can be approximated by a power law, $n_s(\Delta r)=n_0\Delta r^{\alpha}$, on certain ranges. A uniform or random distribution would produce an index $\alpha=0.0$ \citep{Wal84,Gwi94a,Ima02}.
 
The two-point spatial correlation function allows us to find a characteristic scale size for clusters of maser spots as well as the statistical properties of the clustering of spots.   

\subsection{Velocity correlation functions}
Velocity correlation functions measure the variation in Doppler velocity with spot separation. We have used the velocity correlation functions defined by \citet{Wal84}, \citet{Gwi94a}, and \citet{Ima02} in their statistical studies. 

The first one is the rms Doppler-velocity difference as a function of spot separation ($\Delta r$) on the plane of sky. It is expressed by
\begin{equation}
V_s(\Delta r)=\Bigg[\frac{\sum_{i,j}(V_i-V_j)^2n_{\delta}(\vert \boldsymbol{r}_i-\boldsymbol{r}_j\vert)}{\sum_{i,j}n_{\delta}(\vert \boldsymbol{r}_i-\boldsymbol{r}_j\vert)}\Bigg]^{1/2}\,,
\end{equation}
where $n_{\delta}(r)$ is given by equation (2).
The second one is the median Doppler-velocity difference of the spots as a function of $\Delta r$. It can be written as
\begin{equation}
\begin{array}{lcl}
M_s(\Delta r)& = & \mathrm{Med}\,(\vert V_i-V_j\vert) \quad\textrm{for}~i,~j\\
             &   & \textrm{so that}\quad\Delta r < \vert \boldsymbol{r}_i-\boldsymbol{r}_j\vert < \Delta r+dr\,. 
\end{array}
\end{equation}
Again, the indices $i$, $j$ run over all spots, $\Delta r=(\Delta x^2+\Delta y^2)^{1/2}$, and $dr$ is the separation of the adjacent bins in the $V_s$ and $M_s$ vs. $\Delta r$ plots. Generally, these velocity correlation functions can be fitted by a power law ($\Delta r^{\alpha}$) in the whole range. The $M_s$ function  is less sensitive to outlying points with a large velocity difference \citep{Gwi94a}.

Velocity correlation functions give information about the gas kinematics traced by maser spots. 

\section{Star-Forming Regions: Cepheus A and W75 N}
Cepheus A is a high mass SFR located at a distance of $\simeq$700~pc \citep{Mos09}. The brightest radio continuum source of the region, HW2, is a thermal radio jet \citep{Hug84,Rod94,Gom99}. This source is associated with a cluster of water masers detected by \citet{Tor96} using the Very Large Array (VLA). Through VLBA multi-epoch water maser observations toward the radio jet HW2, \citet{Tor01b} detected $\simeq$1000 water maser spots in each of the three observed epochs, forming remarkable linear/arcuate structures with sizes of the order of tens of AU. These structures are grouped in five subregions (named as R1 to R5) within $\sim$0\farcs6 ($\sim$400~AU) from HW2. The most interesting structure is an arc of $\sim$0\farcs1 size ($\sim$70~AU) located in R5. The maser spot positions can be fitted extremely well by a circle of $\sim$62~AU radius \citep[with a precision of 0.1\%,][]{Tor01a}, and their proper motions indicate uniform expansion of $\sim$9~km~s$^{-1}$ perpendicular to the arc. Both the spatial distribution and the proper-motion direction of water masers strongly suggest that the arc is part of a spherical expanding structure excited by a YSO, probably located $\sim$0\farcs6 ($\sim$400~AU) south of HW2 \citep{Cur02}.

W75 N is also a high mass SFR located at a distance of $\simeq$2~kpc \citep{Dic69}. At scales of arcseconds (few thousands of AUs), there are three ultracompact radio continuum sources, VLA~1, VLA~2, and VLA~3 \citep{Hun94,Tor97}, which are probably excited by early B-type stars \citep{She03,She04,Per06}. VLA observations toward this region revealed two clusters of water masers, one associated with VLA~1 and the other associated with VLA~2 \citep{Tor97}. In order to study in detail not only the spatial distribution of water masers but also the gas kinematics in these sources, \citet{Tor03} carried out VLBA multi-epoch water maser observations to measure their proper motions. They detected $\sim$700 maser spots in each of the three observed epochs, which were distributed in two groups, each one associated with a radio continuum source, but tracing outflows with clearly different characteristics. In VLA~1 the maser spots form a linear structure of $\sim$0\farcs75 ($\sim$1500~AU) along the major axis of the radio continuum emission with a mean proper-motion value of $\sim$19~km~s$^{-1}$ parallel to this major axis. On the other hand, in VLA~2 the maser spots describe a shell of $\sim$0\farcs16 size ($\sim$320~AU) around the radio continuum emission. The masers move outward from the central source in multiple directions with a mean proper-motion value of $\sim$28~km~s$^{-1}$. The degree of collimation is clearly different in the two outflows traced by the water masers in this SFR.

The largest maser structures of these two SFRs are located in subregions R5 in Cepheus~A, and VLA~1 and VLA~2 in W75~N. Each structure is associated with a YSO and could give fundamental information about the gas dynamics very close to the protostar. Since the number of maser components is very large, it is important to do statistical studies of the properties of the spatio-kinematical distribution of these masers from scales smaller than 1~AU up to a few hundred of AUs.

\section{Results}
\subsection{Statistical properties of the spatial distribution of water masers}
Using VLBA data of three epochs of water maser observations toward the SFRs Cepheus A and W75 N \citep{Tor01b,Tor03}, we compute two-point spatial correlation functions of maser spots in the following subregions: R5 in Cepheus A, and VLA 1 and VLA 2 in W75 N. In each of these subregions, a few hundred maser spots were detected with an S/N $\geq$ 10 in each one of the three epochs (see Table 1). The uncertainties in the values of these functions are mainly due to position uncertainty of maser spots, which is more significant toward small separations (represented by the error bar sizes shown in Figures 1-3). We estimated the sizes of the error bars in the two-point spatial correlation functions in the following way: starting from the observed spatial distribution of maser spots, we have applied random displacements, following a Gaussian probability distribution, with a dispersion equal to the positional uncertainty. We have calculated several such alternative distributions of masers and estimated the correlation function of each of them. In this way, we obtained a set of different alternative correlation functions. Then, we calculated the standard deviation of the set of values of the correlation function at each separation and adopted that deviation as the typical error of the correlation function.

We approximate the two-point spatial correlation functions by two different power laws, one within small separations and the other within larger separations, because there is a hint of a break in the slope of the correlation function. An approximation with two different power laws has a smaller value of chi-square than a single power-law fit. We calculate the best value for the breakpoint as that for which the chi-square in the fit is minimum.
  
We have adopted a minimum value of the scale range of 0.07~AU for Cep~A R5, and of 0.2~AU for VLA~1 and VLA~2 in W75 N, corresponding to two times the accuracy in relative positions of maser spots (2$\sigma\simeq0.1$~mas) for each case. At separations smaller than this value, the correlation function is not accurate, since positional information is dominated by random noise. The power-law indices and the scale ranges of the fits are listed in Table 2.

Two-point correlation functions for maser spots in Cep A R5 for the three observed epochs are shown in Figure 1.
These functions follow a power-law dependence between 0.07 and 0.24~AU with a weighted mean value of the index $\alpha=-2.79\pm0.22$ (Table 2), illustrated as a straight line in this figure. At a separation of 0.31$\pm$0.07~AU, there is a break of the power-law slope (indicated by shades). A different power-law dependence is found in the scale range of 0.5$-$70~AU, with a weighted mean value of the index, $\alpha=-1.46\pm0.09$ (Table 2).

Figure 2 shows the two-point correlation functions for VLA~1 in W75~N. In VLA 1, the number of maser spots detected is smaller than in the other regions, introducing larger errors in our statistical analysis. In fact, we have found that the values of the correlation function at separations smaller than 1 AU are more uncertain. Large error bars are due to both the relative position uncertainty of the maser spots and small number (a few tens) of maser spots at these separations. At separations between 0.2 and 0.75~AU, the correlation function roughly follows a power with a weighted mean value of the index $\alpha=-3.92\pm0.33$ (Table 2). There is a hint of a break of the power-law slope at a separation of 0.85$\pm$0.10~AU (indicated by shades). Another power law is found in the scale range of 1.3$-$1500 AU, with a weighted mean value of the index, $\alpha=-1.47\pm0.10$.

Figure 3 shows the two-point correlation functions for VLA~2 in W75~N. These functions follow a power law between 0.2 and 0.6~AU with a weighted mean value of the index $\alpha=-2.51\pm0.17$ (Table 2). There is a break of the power-law slope at a separation of 0.9$\pm$0.3~AU (indicated by shades). A different power law is found in the scale range of 1.5$-$250 AU, with a weighted mean value of the index, $\alpha=-1.30\pm0.07$.

\subsection{Statistical properties of the velocity distribution of water masers}
We have calculated the velocity correlation functions defined in \S 2.2 for water maser spots in the subregions R5 in Cepheus A, and VLA~1 and VLA~2 in W75~N. The velocity resolution was 0.21~km~s$^{-1}$ and 0.4~km~s$^{-1}$ for the Cepheus A and W75~N observations, respectively.

In general, both velocity correlation functions, $V_s(\Delta r)$ and $M_s(\Delta r)$, can be approximated by a power law in the whole range. The power-law indices and the scale ranges of the fits are given in Table 3. These two velocity correlation functions, the rms and the median of the difference in Doppler velocity as a function of the separation between spots, are shown in Figure 4 for subregion Cep A R5. The obtained power-law indices for both functions are consistent with zero for the three observed epochs, with weighted mean values $\alpha=-0.03\pm0.03$ and $\alpha=0.00\pm0.04$ for the rms and the median Doppler-velocity difference, respectively (Table 3). The error bars in the velocity correlation functions are due to both the uncertainty in the relative positions of the maser spots and the uncertainty in their velocities. These error bars were estimated in an analogous way to the two-point spatial correlation function, but including the velocity uncertainty, i.e., letting the velocity of the maser spots vary within the velocity uncertainty (the channel width in the observations). Then we estimate the velocity correlation function of each new distribution and compare the results with the original, estimating a typical error from the standard deviation of the velocity correlation function after several trials.
 
Figures 5 and 6 show the velocity correlation functions for VLA~1 and VLA~2 in W75~N. In these subregions, the correlation functions show a trend of increasing Doppler-velocity difference with increasing separation. If these functions are approximated by a power law, in VLA~1 the weighted mean value of power-law indices are $\alpha=0.17\pm0.05$ and $\alpha=0.14\pm0.05$ for the rms and the median Doppler-velocity difference, respectively (Table 3). In VLA~2, the weighted mean value of the power-law indices are $\alpha=0.13\pm0.03$ and $\alpha=0.20\pm0.02$ for the rms and the median Doppler-velocity difference, respectively.

\section{Discussion}
\subsection{Clustering}
In order to understand the information given by the two-point spatial correlation functions of maser spots in these SFRs, we carried out a similar statistical analysis for some hypothetical cases with an illustrative purpose. For example, we calculated the two-point spatial correlation function for a cluster of $\sim$10000 dots randomly and uniformly distributed in a circle with a diameter equal to 1.0 (in arbitrary units). The results are shown in the first row of Figure 7. At separations much smaller than 1, the curve has an index $\alpha=0.0$ because the number of dots per unit area does not depend on dot separation. The curve drops rapidly at separations close to the diameter of the circle, as a result of the fact that the dots are confined to a finite region.

We have also studied the case of a discrete circle with a diameter equal to 1.0 formed by several circular clusters, each one with a diameter equal to 0.15. In the case of three clusters, the distance between any pair of clusters is the same (see Figure 7, second row). The two-point correlation function for such a distribution shows that the number of dots per unit area decreases rapidly at separations close to the diameter of each cluster. As separation becomes larger than 0.15, the number of dots per unit area increases gradually, because we start to count dots that belong to another cluster, having a maximum at a separation equals to the value of the distance between the centers of the clusters. Increasing the number of clusters results in an increase of the number of maxima, at separations equal to the different values of the distance between the centers of the clusters (see Figure 7, third and fourth rows).
 
From the hypothetical cases described above, we know that the two-point spatial correlation function of a random and uniform distribution of dots can be described by a power-law dependence with an index $\alpha=0.0$. Therefore, the power-law indices found for the maser spots in the SFRs Cepheus A and W75 N (see Table 2) strongly suggest a characteristic clustering of maser spots. This behavior indicates that the water maser spots are associated in a scale-free or fractal fashion on the sky, as noted for other SFRs, such as W49N, W3 IRS 5, and Sgr B2(M) \citep{Gwi94a,Ima02,Str02}.

From Figure 7, we can estimate the characteristic size of the clusters of dots as the scale or separation where the first steep drop of the two-point correlation function occurs. Inspection of Figures 1, 2, and 3 reveals a break in the slope of the two-point correlation functions of the three studied regions (Cep A R5, W75~N VLA~1, and W75~N VLA~2) that is highlighted by shades. If we identify this breakpoint as the characteristic scale, we can estimate a typical size for clusters of maser spots of $\sim$0.31$\pm$0.07~AU in Cep A R5. This value is slightly smaller than the value found in W3 IRS 5 \citep[$\sim$0.5~AU,][]{Ima02}. For VLA~1 and VLA~2 in W75~N, the typical size for clusters of maser spots is $\sim$0.85$\pm$0.10~AU, and $\sim$0.9$\pm$0.3~AU, respectively. These values found in W75 N are very close to the typical size of water maser features of 1~AU found in the SFR W49 N \citep{Gwi94a}. We note that the typical size for clusters of water maser spots tends to be $\la$1 AU in all these SFRs. This may be related with a scale for water maser excitation. Theoretical models of water masers based on shock excitation yield to a similar scale size for masing regions \citep{Eli89,Kau96}. 

Additionally, we note that in the regions studied in this work, the typical size of the clusters of maser spots is smaller in Cepheus A than in W75 N. This could reflect a systematic error (for instance, in distance determination), or be an intrinsic property of the regions (depending on their evolutionary stage or stellar mass, for instance). Studies in other regions with the tools used in this paper would be necessary to identify whether these differences have a physical meaning.

At separations larger than the characteristic scale of clusters, the two-point correlation functions of the modeled distribution of clusters show oscillations with local maxima located at separations equal to the values of the distance between the centers of the clusters (see Figure 7). If we observe in detail the close-up of the two-point correlation functions of maser spots in the three observed regions (Figs. 1, 2, and 3), we then note these oscillations at separations larger than the breakpoint. The local maxima may correspond to the typical distances ($\sim$0.7~AU in Cep A R5, and $\sim$3~AU in W75~N VLA~1 and W75~N VLA~2) between clusters of maser spots.

The change of the power-law index between small scales ($\alpha=-4.3$ to $-2.4$) and large scales ($\alpha=-1.56$ to $-1.27$) mentioned in \S 4.1, suggests that the statistical properties of the clustering of maser spots are different at scales smaller and larger than the typical size of clusters.

Furthermore, we have found that the value of the index $\alpha$ hardly changes from one epoch to another, at separations larger than the characteristic scale size for clusters of water masers (Table 2). This behavior suggests that the spatial distribution of clusters of water masers does not change significantly on large scales during the three observed epochs. In fact, structures of several tens of AU are preserved during the time span of the observations (two months). On the other hand, the value of the index $\alpha$ apparently changes from one epoch to another at separations smaller than the characteristic scale size for clusters (Table 2). This probably indicates a change in the density or spatial distribution of the maser spots inside each cluster.

\subsection{Velocity distribution}
\citet{Str02} investigated whether regular (non-turbulent) velocity fields, such as expansion and/or rotation can produce the power-law dependence of the velocity increments on spatial scales, described in the velocity correlation functions. Specifically, they used a model of 90 dots randomly and uniformly distributed in a thin spherical shell considering radial expansion, rotation around an axis perpendicular to the line of sight, and both motions. The line-of-sight velocity difference as a function of dot separation can be fitted by a power law in a given range \citep[see Fig. 8 of][]{Str02}, with an index of about zero for the case of expansion, changing to approximately unity for the case of rotation. When both motions are present, the index has intermediate values. Based on these results, for a spherical geometry, we can conclude that if organized motion (expansion and/or rotation) dominates over turbulence over a certain scale range, the value of the index is expected to be between 0.0 and 1.0. 

In the Cep A R5 and W75~N VLA~2 regions, water masers delineate remarkable arc structures, showing proper motions with a mean value of $\sim$9~km~s$^{-1}$ and $\sim$28~km~s$^{-1}$, respectively (see \S 3). The spatial distribution of the masers and the direction of the proper motions are consistent with expanding motions in a shell \citep{Tor01b,Tor03}. On the other hand, in the  W75~N VLA~1 region, water masers form a linear structure with a mean proper-motion value of $\sim$19~km~s$^{-1}$ along the direction of the elongated structure that was interpreted as tracing a collimated outflow \citep{Tor03}.
       
In our statistical analysis of the velocity distribution of the water masers in Cep~A R5, we find that the velocity correlation functions of both the rms and the median Doppler-velocity difference can be fitted by a power law with an index of about zero for the three observed epochs (Figure 4). This index is consistent with the case of pure expansion of dots randomly distributed in a thin spherical shell \citep{Str02}. This result is in agreement with the VLBA proper-motion measurements of water masers, which indicate expansion of a spherical shell \citep{Tor01a,Tor01b}.

On the other hand, the value of the power-law index found for the velocity correlation functions of maser spots in W75~N VLA~2 is $\sim$0.18, suggesting another geometry (different from a spherical shell) for this source. In fact, in W75~N VLA~2, masers delineate an incomplete elliptical ring on the plane of the sky; while in Cep A R5, masers delineate part of a circle on the plane of the sky (a circle in projection is produced only by a limb-brightened spherical shell). In addition, the masers in W75~N VLA~2 present a higher Doppler-velocity dispersion than maser spots in Cep A R5 \citep[25~km~s$^{-1}$ vs. 2~km~s$^{-1}$;][]{Tor01b,Tor03}.

For the reasons described above, we have considered a ring with an arbitrary orientation with respect to the observer, applying a similar statistical analysis to a model that simulates organized motions of masers. Our aim is to determine whether a particular orientation and the presence of expansion and/or rotation motions could explain the trend of increasing Doppler-velocity difference with increasing separation of spots indicated by the value of the power-law index of the velocity correlation function in W75~N VLA~2 (Fig. 6 and Table 3).

We have used $\sim$1000 dots randomly and uniformly distributed in a ring with an outer radius equal to $\sim$0\farcs08 ($\sim$160~AU), the radius of the region traced by masers in W75~N VLA~2. The inner radius of the ring is chosen in such a way that the difference between the outer and inner radii is equal to $\sim$0\farcs02 ($\sim$40~AU), the spatial dispersion of the masers observed within the VLA~2 shell. Also, the vertical thickness of the ring is assumed to have the same value. We have analyzed three types of organized motions: uniform expansion, rigid-body rotation around an axis with an inclination angle $\theta$ with respect to the line of sight, and expansion plus rotation. In our calculations, the magnitudes of the expansion and rotation velocities are assumed to be equal. In order to estimate the uncertainties involved in these calculations, we have considered that each different spatial configuration of the dots randomly distributed within the ring would result in slightly different velocity correlation functions. Therefore for each type of organized motion, we have analyzed different spatial configurations of dots to find the mean values of the $V_s$ function and its dispersion. The results of the rms line-of-sight velocity difference ($V_s$) as a function of dot separation are qualitatively similar to the median line-of-sight velocity difference ($M_s$); therefore, we only present the $V_s$ function.

The mean values of ($V_s$) as a function of dot separation are shown in the top panel of Figure 8, presenting a comparison between the three types of motions for a ring observed with an inclination angle of $\theta=20\degr$ (the angle between the line of sight and the normal to the ring plane; $\theta=0\degr$ corresponds to the face-on case). The magnitudes of both the expansion and rotation velocities are equal to $\sim$44~km~s$^{-1}$. These values of the inclination angle and velocities yield the best fit of the data. For the case of expansion (black dotted curve), the $V_s$ function is almost constant for dot separations smaller than $\sim$0\farcs01 ($\sim$20~AU), and increases smoothly for larger dot separations. This trend is not well approximated by a power-law dependence. 
For the case of rotation (black dashed curve), $V_s$ increases continuously with dot separation. This tendency can be satisfactorily approximated by a power-law dependence with an index of about 1. For the case of expansion plus rotation (black continuous curve), $V_s$ is similar to that of the pure expansion case at separations smaller than $\sim$0\farcs01 ($\sim$20~AU), however when including rotation, $V_s$ increases more rapidly for larger dot separations.

In the middle panel of Figure 8, we show a comparison between the velocity correlation functions calculated for different spatial configurations of dots randomly and uniformly distributed in an arc, equal to one half of the previously studied ring with the same orientation with respect to the observer, and with equal values of expansion and rotation velocities. The selected section of the ring is predominantly redshifted. The values of $V_s$ as a function of dot separation are similar to those for a complete ring for separations $\lesssim$0\farcs05 ($\lesssim$100~AU); at larger separations, $V_s$ decreases slightly with respect to the complete ring case, because in the case of a half ring the maximum velocity difference is smaller. Since maser spots in W75 N VLA~2 do not trace a continuous or complete structure on the plane of the sky, we consider that this scenario might help us to understand the decrease of $V_s$ observed in the velocity correlation functions on a scale larger than $\sim$100~AU (see Figure 6). 

In fact, the bottom panel of Figure 8 shows the rms values of Doppler-velocity difference ($V_s$) calculated for the masers in W75~N VLA~2 observed in the second epoch (when more maser spots were detected), together with the modeled functions that best fit these data (shown in the middle panel). Between 1 and 100 AU (vertical dotted lines), there is a general agreement between the functions calculated for the modeled spots and the observed masers. For the modeled functions, considering pure expansion or expansion plus rotation produce similar results in this scale range. However, at separations larger than 100 AU, the $V_s$ function of the modeled spots decreases more rapidly when rotation motions are included, which is qualitatively similar to the $V_s$ of the observed masers at those separations. At separations $<$1~AU, a larger dispersion is present in the velocity correlation functions for the modeled spots (shown by the grey curves), due to a lack of clustering, and thus poor statistics. This occurs because even though the total number of modeled spots is large, the number of pairs with small separation is small because modeled spots do not form clusters (they are uniformly and randomly distributed).

From our results, we thus conclude that the trend described by the rms Doppler-velocity difference of maser spot pairs in W75 N VLA~2 as a function of separation can be explained by the presence of organized motions, expansion, and rotation (both velocities with the same magnitude $\sim$44~km~s$^{-1}$) within an arc of a circle observed with small inclination angle, $\theta\simeq$20\degr. Higher values of the angle $\theta$ will produce a steeper gradient of the velocity difference toward larger separations. Moreover, the corresponding projected ellipse on the plane of the sky would have its minor axis much smaller than its major axis, which is not observed. The expansion velocity required to fit the data is about the same order of magnitude as the measured proper motions \citep{Tor03}.

The source W75 N VLA~1 appears to be different from VLA~2 since proper-motion measurements of water masers indicate a collimated outflow in this region \citep{Tor03}. In fact, the velocity correlation functions of maser spots present a larger dispersion from the fit. Despite the dispersion, the values estimated for the indices (similar to those of VLA~2) suggest the presence of an organized component of motion with a preferential direction.  

\section{Conclusions}
In this paper, we present a statistical analysis of the spatial and velocity distribution of water masers in the SFRs Cepheus A and W75~N, using the data of the VLBA maser observations carried out by \citet{Tor01b,Tor03}. 
Our conclusions are as follows:

\begin{enumerate}
\item We have found a characteristic scale size for clusters of water maser spots $\la$1~AU, indicated by the break in the slope of the two-point spatial correlation function. Specifically, $\sim$0.31$\pm$0.07~AU in Cep A R5, $\sim$0.85$\pm$0.10 in W75~N VLA 1, and $\sim$0.9$\pm$0.3~AU in W75~N VLA2. These values are close to the typical sizes of the water maser features found in other SFRs, e.g., $\sim$0.5~AU in W3~IRS~5 \citep{Ima02} and $\sim$1~AU in W49~N \citep{Gwi94a}. Probably, these results may indicate that the scale for water maser excitation tends to be $\la$1 AU, as it was pointed out by models of water masers excited by shocks \citep{Eli89,Kau96}.

\item Two-point spatial correlation functions follow power-law dependences, indicating self-similar spatial distributions of water masers. The power-law indices found for separations smaller than the characteristic scale size of clusters ($\alpha=-$4.3 to $-$2.4) are steeper in Cepheus A and W75~N than the value found in W3 IRS~5 \citep[$\alpha=-2.09$,][]{Ima02}. This could be due to a different spatial distribution of spots inside the clusters of each region. The power-law indices found for separations larger than the characteristic scale size ($\alpha=-1.56$ to $-$1.27) are similar to the value found in W49~N \citep[$\alpha=-1.33$,][] {Gwi94a}. This suggests similar statistical properties for the spatial distribution of masers at scales larger than 1 AU in the SFRs Cepheus A, W75~N, and W49~N.

\item Velocity correlation functions follow power-law dependences. In Cep~A R5, the power-law index is about zero, consistent with pure expansion of dots randomly distributed in a thin spherical shell. In W75~N VLA~2, the value for the power-law index can be explained by the presence of expansion and rotation within an arc of a circle observed with a small inclination angle. These results are in agreement with proper-motion observations of water masers \citep{Tor01a,Tor01b,Tor03}. In W75~N VLA~1, the value for the power-law index seems to be consistent with organized motions but further studies are needed. In the SFRs Cepheus A and W75~N, the values estimated for the power-law indices suggest that water masers are tracing an organized component of the velocity field instead of a turbulent component, probably due to a prevalence of organized motions over turbulent motions at spatial scales from a few hundred of AUs down to less than 1~AU.

\item Based on the value of the power-law index followed by the velocity differences of maser spot pairs as a function of their separation, it may be possible to find evidence for expansion and/or rotation motions by analyzing water maser data from a single observed epoch.
\end{enumerate}

\acknowledgments
L.~U. is supported by Secretar\'{\i}a de Estado de Universidades e Investigaci\'on of MEC (Spain). G.~A., J.~F.~G., J.~M.~T., and L.~U. are partially supported by Ministerio de Ciencia e Innovaci\'on (Spain), grant AYA 2008-06189-C03 (including FEDER funds), and by Consejer\'{\i}a de Innovaci\'on, Ciencia y Empresa of Junta de  Andaluc\'{\i}a (Spain). J.~C., A.~C.~R., and S.~C. acknowledge the support of CONACyT (Mexico) grants 61547 and 60581. We are thankful to our referee for his/her useful comments that help us to improve this paper.

\clearpage

\begin{figure}
\epsscale{0.9}
\plotone{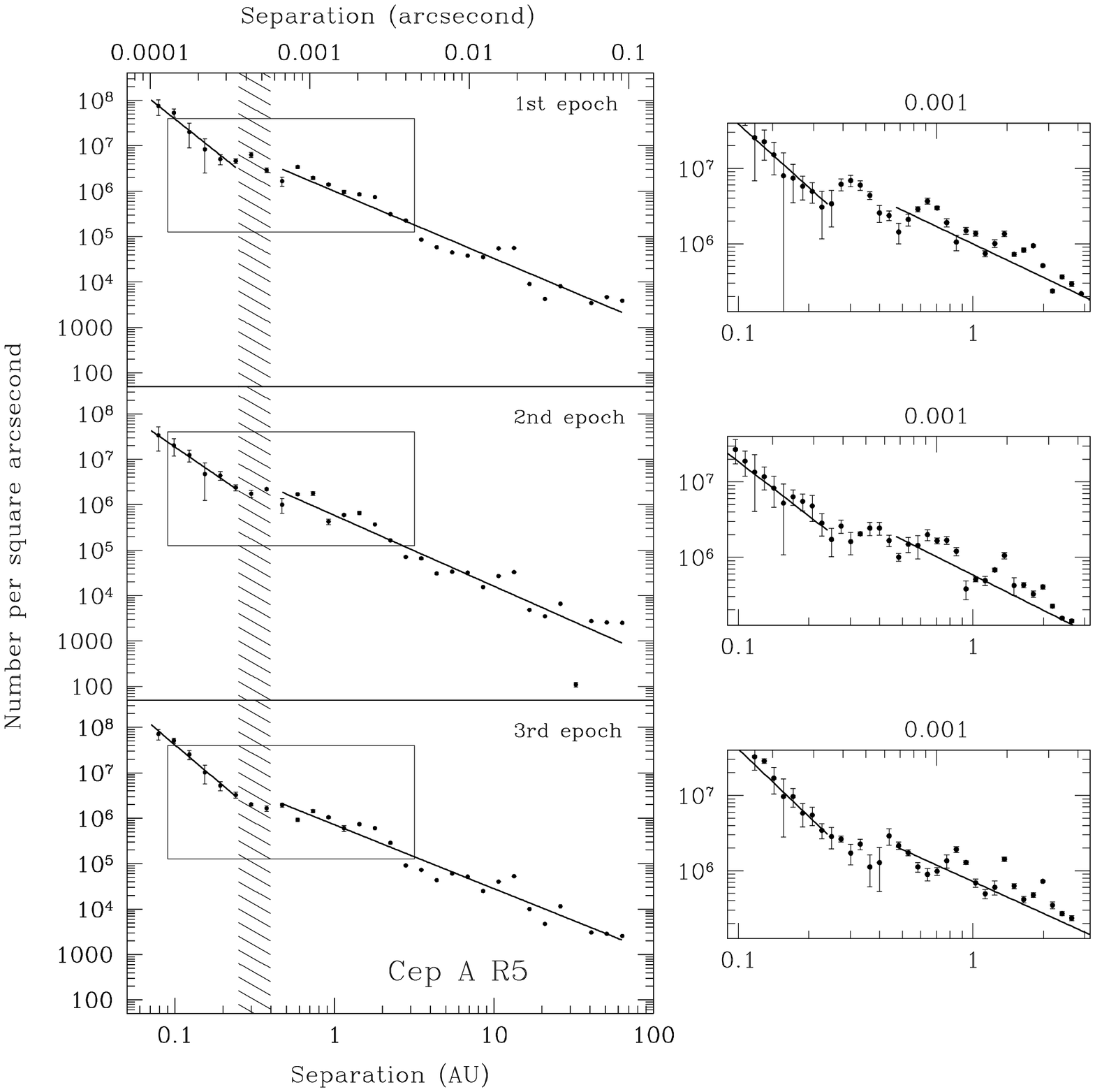}
\caption{\textit{Left panels:} Two-point correlation functions for water maser spots in subregion Cep A R5 for each of the three observed epochs. The straight lines show power-law fits in two different scale ranges, 0.07$-$0.24~AU and 0.5$-$70~AU (see Table 2). The shaded region shows the scale range within which the power-law slope changes. The positional accuracy (2$\sigma$) of the spots is 0.07 AU. The error bars are mainly due to position uncertainty of maser spots.
\textit{Right panels:} Close-up of the correlation function for each of the three observed epochs. We recomputed this function using a smaller value of $dr$ to see in detail the break of the power-law slope. The lines show the same power-law fits as in the left panels.
 \label{fig1}}
\end{figure}

\clearpage
\begin{figure}
\epsscale{0.9}
\plotone{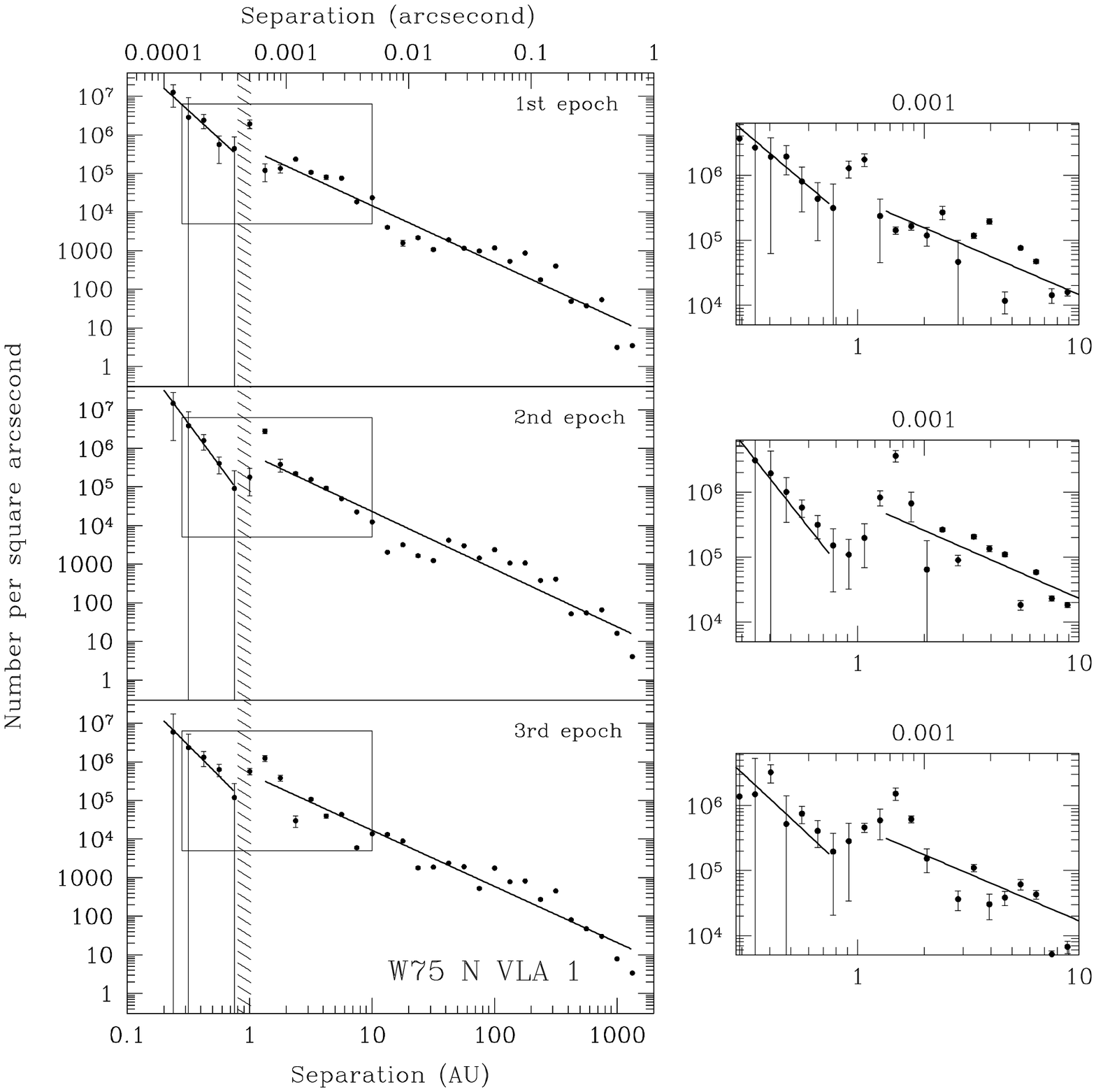}
\caption{\textit{Left panels:} Two-point correlation functions for water maser spots in subregion W75 N VLA 1 for each of the three observed epochs. The straight lines show power-law fits in two different scale ranges, 0.2$-$0.75~AU and 1.3$-$1500~AU (see Table 2). The shaded region shows the scale range within which the power-law slope changes. The positional accuracy (2$\sigma$) of the spots is 0.2 AU. \textit{Right panels:} Close-up of the correlation function for each of the three observed epochs. We recomputed this function using a smaller value of $dr$ to see in detail the break of the power-law slope. The lines show the same power-law fits as in the left panels.
 \label{fig2}}
\end{figure}

\clearpage

\begin{figure}
\epsscale{0.9}
\plotone{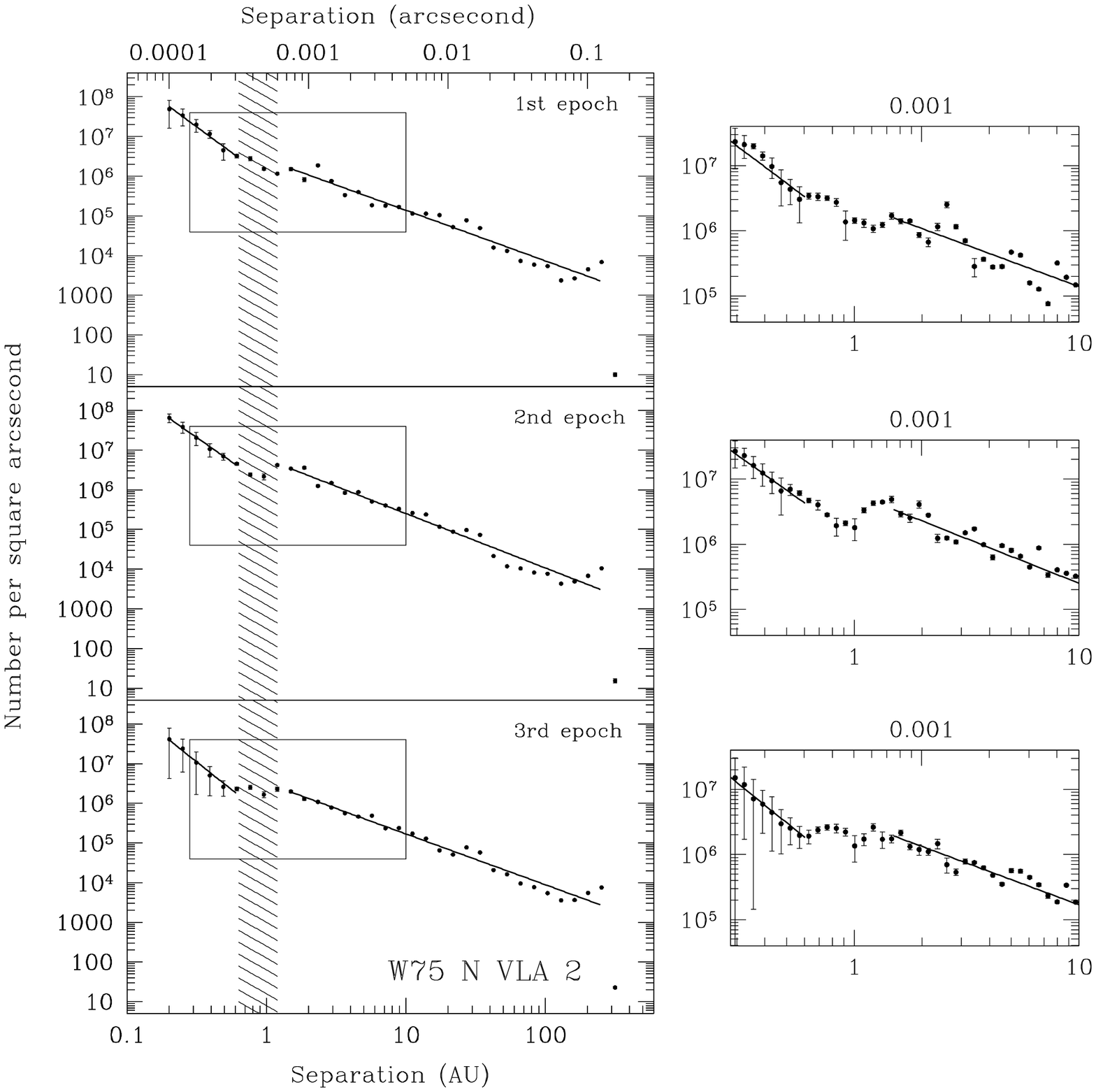}
\caption{\textit{Left panels:} Two-point correlation functions for water maser spots in subregion W75 N VLA 2 for each of the three observed epochs. The straight lines show power-law fits in two different scale ranges, 0.2$-$0.6~AU and 1.5$-$250~AU (see Table 2). The shaded region shows the scale range within which the power-law slope changes. The positional accuracy (2$\sigma$) of the spots is 0.2 AU. \textit{Right panels:} Close-up of the correlation function for each of the three observed epochs. We recomputed this function using a smaller value of $dr$ to see in detail the break of the power-law slope. The lines show the same power-law fits as in the left panels.
 \label{fig3}}
\end{figure}

\clearpage

\begin{figure}
\epsscale{0.9}
\plotone{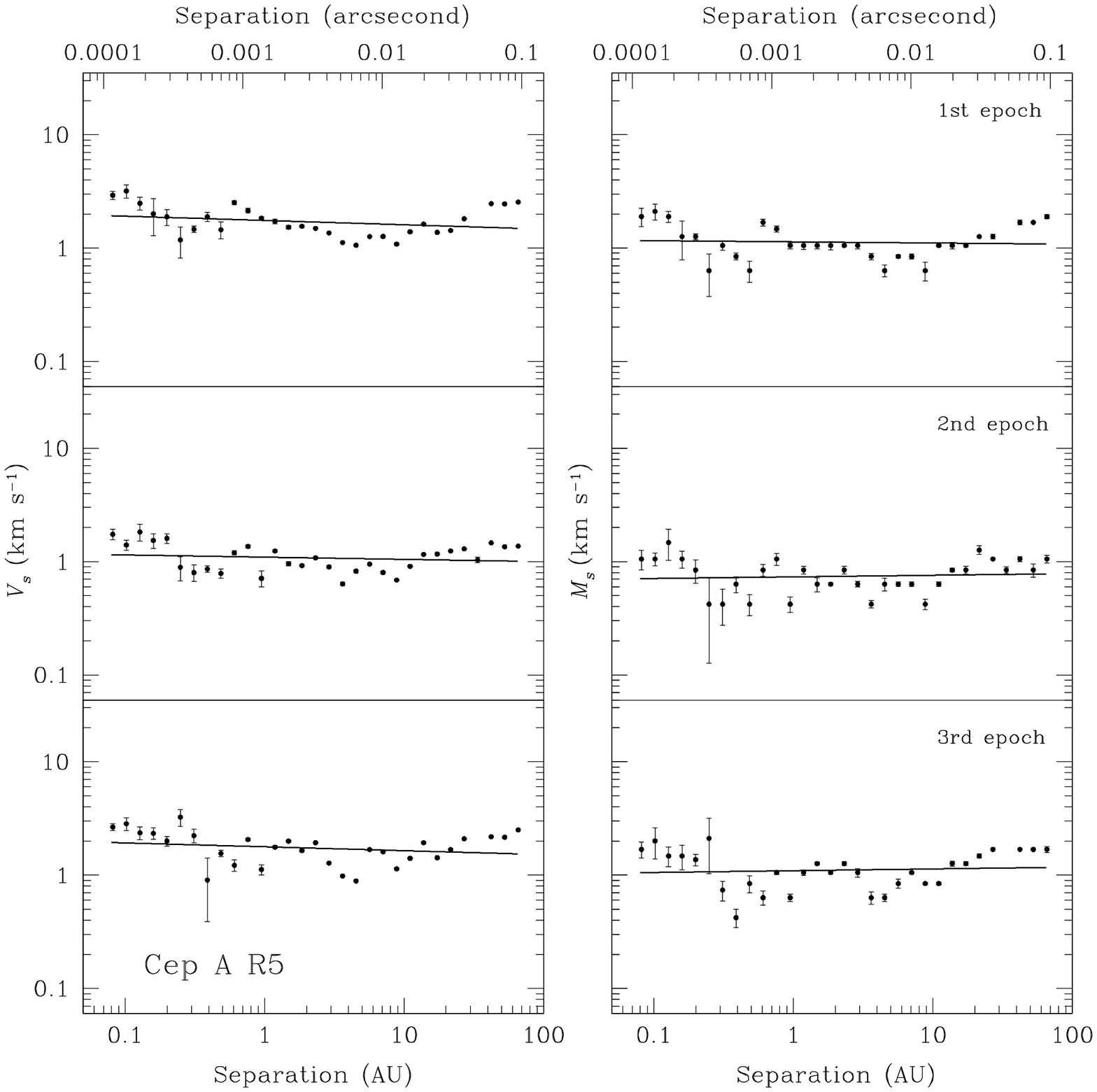}
\caption{Velocity correlation functions for water maser spots in Cep A R5 for each of the three observed epochs. Left panels show the rms values of Doppler-velocity differences ($V_s$) between maser spot pairs as a function of their separation on the plane of the sky. Right panels show the medians of the Doppler-velocity differences ($M_s$) between maser spot pairs as a function of their separation on the plane of the sky. The straight lines show power-law fits (Table 3). The positional accuracy (2$\sigma$) of the spots is 0.07 AU. Error bars are due to position and velocity uncertainties of maser spots.
 \label{fig4}}
\end{figure}

\clearpage

\begin{figure}
\epsscale{0.9}
\plotone{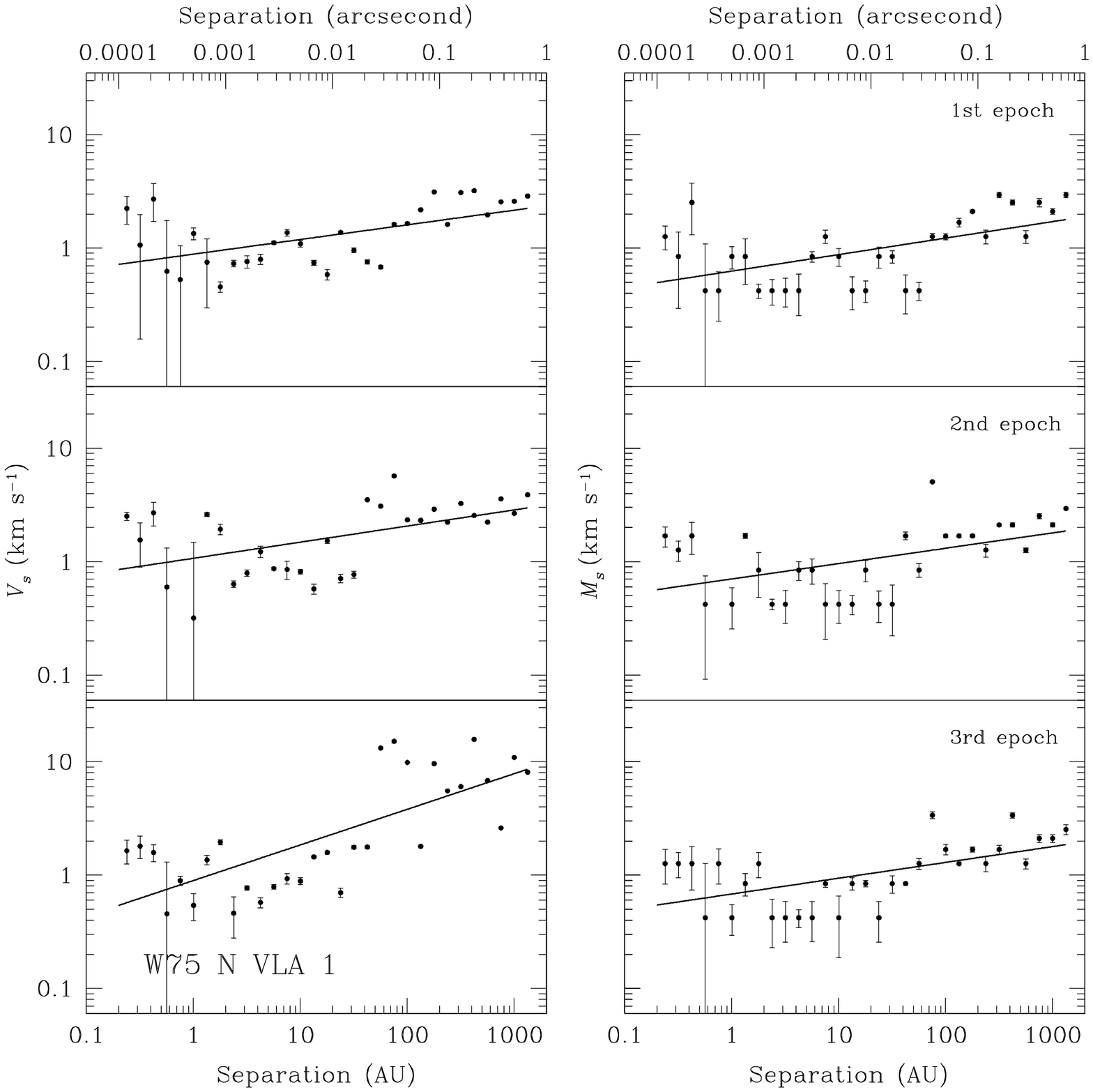}
\caption{Velocity correlation functions for water maser spots in W75 N VLA 1 for each of the three observed epochs. Left panels show the rms values of Doppler-velocity differences ($V_s$) between maser spot pairs as a function of their separation on the plane of the sky. Right panels show the medians of the Doppler-velocity differences ($M_s$) between maser spot pairs as a function of their separation on the plane of the sky. The straight lines show power-law fits (Table 3). The positional accuracy (2$\sigma$) of the spots is 0.2 AU.
 \label{fig5}}
\end{figure}

\clearpage

\begin{figure}
\epsscale{0.9}
\plotone{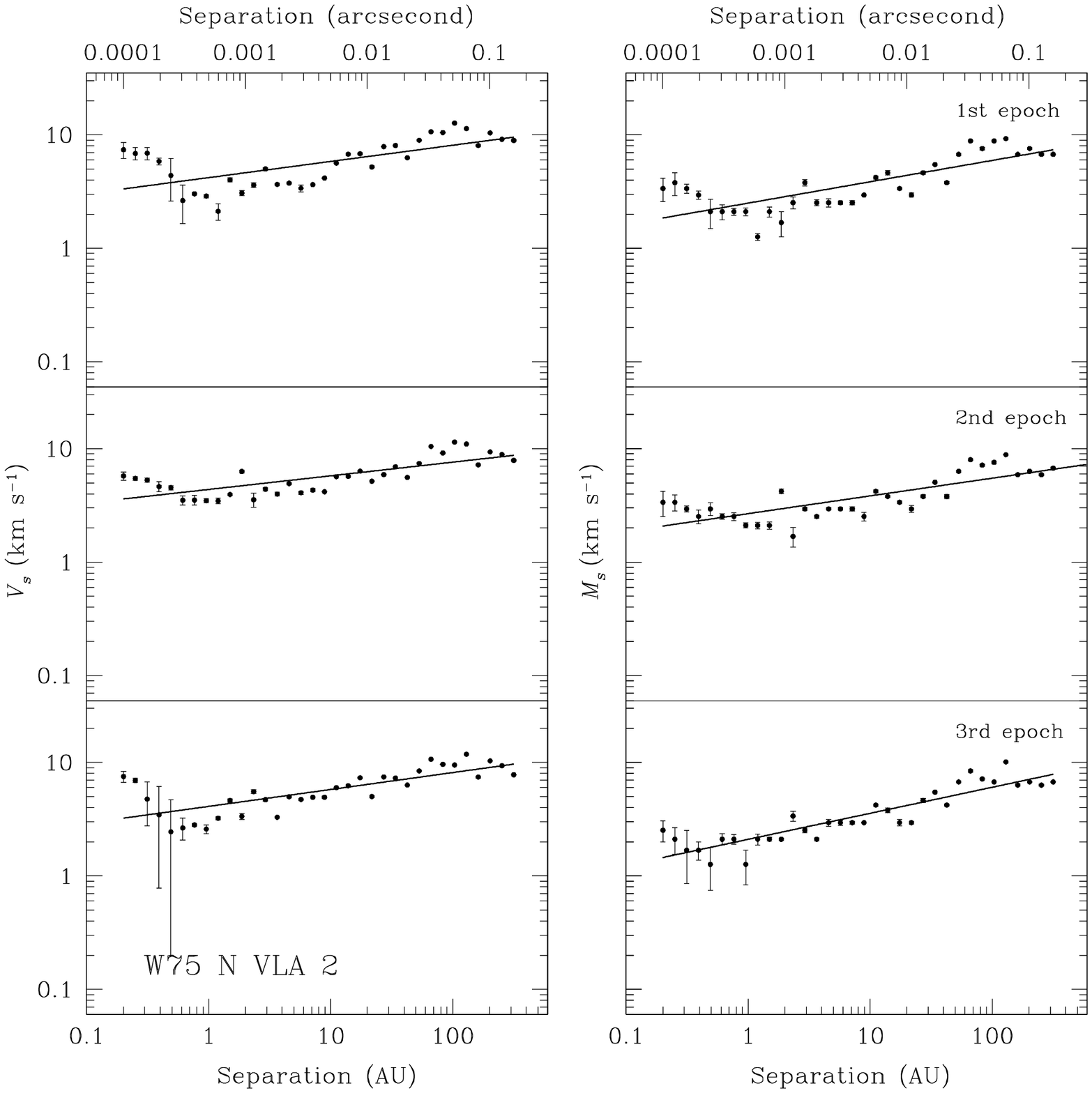}
\caption{Velocity correlation functions for water maser spots in W75 N VLA 2 for each of the three observed epochs. Left panels show the rms values of Doppler-velocity differences ($V_s$) between maser spot pairs as a function of their separation on the plane of the sky. Right panels show the medians of the Doppler-velocity differences ($M_s$) between maser spot pairs as a function of their separation on the plane of the sky. The straight lines show power-law fits (Table 3). The positional accuracy (2$\sigma$) of the spots is 0.2 AU.
 \label{fig6}}
\end{figure}

\clearpage

\begin{figure}
\epsscale{1.1}
\plotone{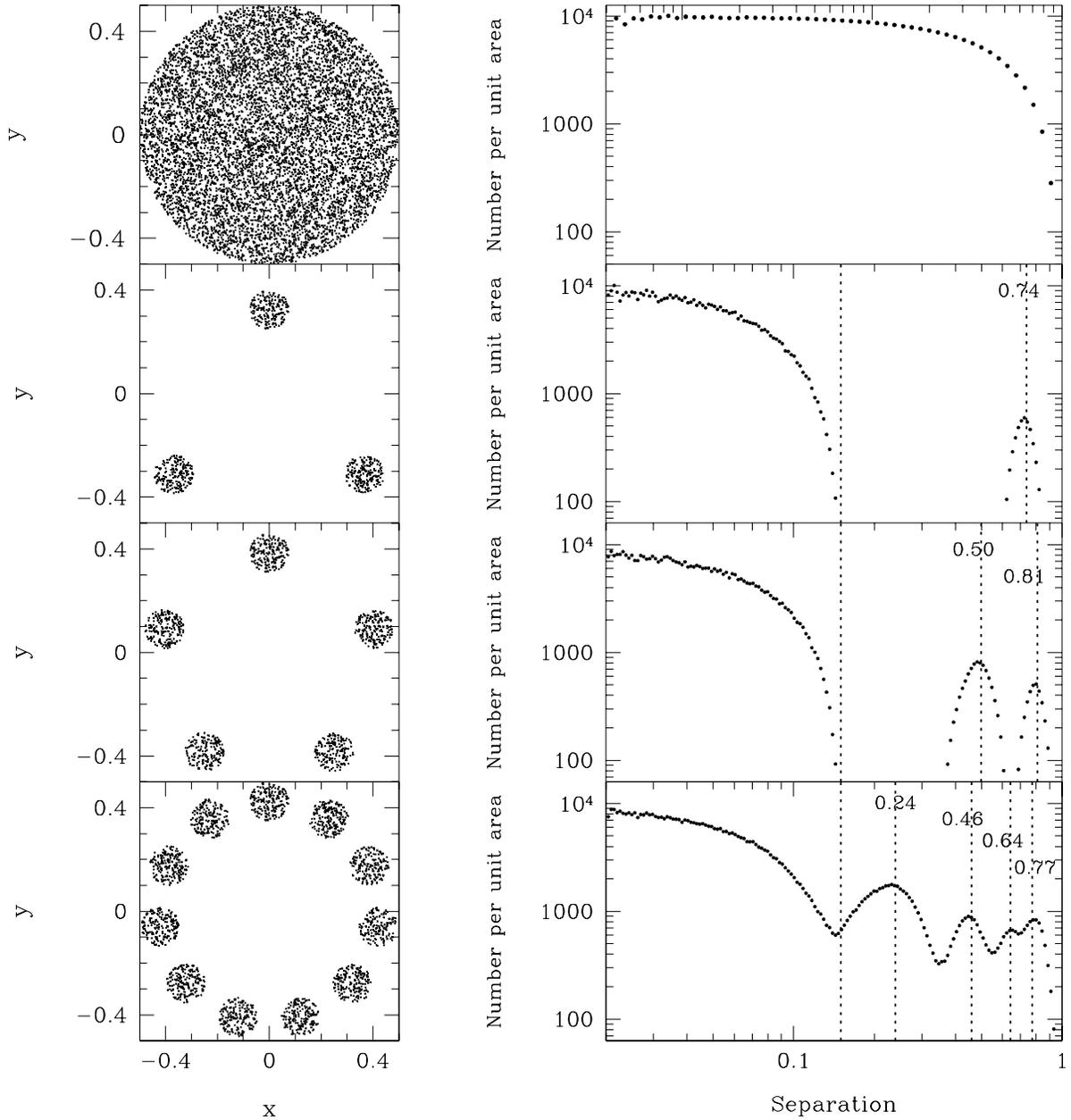}
\caption{
\textit{Left panels} show different distributions of dots and \textit{right panels} show their corresponding two-point correlation functions. From top to bottom, \textit{first row} represents a cluster of $\sim$10000 dots randomly and uniformly distributed in a circle with a diameter equal to 1.0. \textit{Second, third, and fourth rows} represent groups of 3, 5, and 11 circular clusters of dots with a diameter of 0.15, respectively. Dots are randomly distributed within each cluster. Clusters are equally spaced along a circle with diameter equal to 1.0. In right panels, the first dotted line indicates a value of the separation equal to the diameter of a cluster (0.15), and the following dotted lines indicate the distances between the centers of each pair of clusters.
\label{fig7}}
\end{figure}

\clearpage

\begin{figure}
\epsscale{0.9}
\plotone{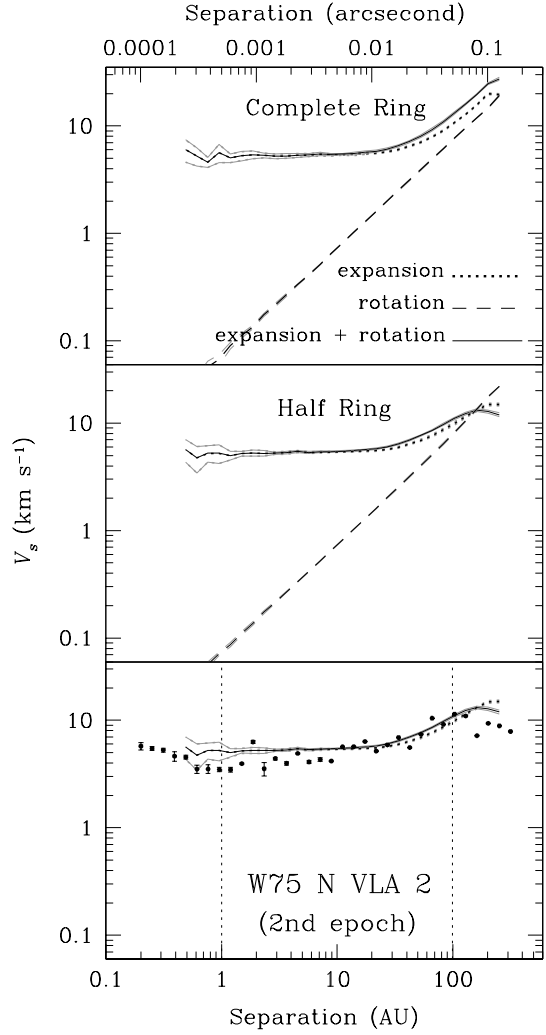}
\caption{\textit{Top}: Velocity correlation functions for $\sim$1000 dots randomly and uniformly distributed in a ring with an inner radius of 120~AU and an outer radius of 160~AU, observed with an inclination angle $\theta=20\degr$. The black dotted, dashed, and continuous curves show the mean values of $V_s$ as a function of separation for models of organized velocity fields: expansion, rotation, and expansion plus rotation. The grey curves show the 1$\sigma$ error associated with these simulations. The magnitudes of both the expansion and rotation velocities are equal to 44~km~s$^{-1}$. The correlation functions are not calculated at separations smaller than 1~AU, because of the increased dispersion (see details in \S 5.2). \textit{Middle}: Same as \textit{top}, but only for one half of the ring using the same kinematical parameters and orientation. \textit{Bottom}: Comparison between the rms values of Doppler-velocity differences ($V_s$) for maser spots in W75~N VLA~2 detected in the second epoch (filled circles) and the velocity correlation functions for pure expansion and expansion plus rotation, that correspond to half a ring, shown in the middle panel.
\label{fig8}}

\end{figure}

\clearpage

\begin{deluxetable}{cc@{\extracolsep{1.5em}}cc@{\extracolsep{1.5em}}cc} 
\tabletypesize{\footnotesize}
\tablecolumns{6} 
\tablewidth{0pc}
\tablecaption{Water maser spots detected in the subregions Cep A R5, W75 N VLA 1, and W75 N VLA 2.} 
\tablehead{  
\multicolumn{2}{c}{Cep A R5} & \multicolumn{2}{c}{W75 N VLA 1} & \multicolumn{2}{c}{W75 N VLA 2}\\
\cline{1-2}\cline{3-4}\cline{5-6}
Epoch & $n_{\mathrm{spot}}$\tablenotemark{a} & Epoch & $n_{\mathrm{spot}}$\tablenotemark{a} & Epoch & $n_{\mathrm{spot}}$\tablenotemark{a}
}
\startdata 
1996 Feb 11 & 283 & 1999 Apr 2 & 126 & 1999 Apr 2 & 521 \\
1996 Mar 10 & 170 & 1999 May 7 & 174 & 1999 May 7 & 791 \\
1996 Apr 13 & 234 & 1999 Jun 4 & 134 & 1999 Jun 4 & 613 \\
\enddata
\tablenotetext{a}{Number of maser spots detected with an S/N $\geq$ 10.}
\end{deluxetable}

\begin{deluxetable}{ccc@{\extracolsep{1.5em}}cc@{\extracolsep{1.5em}}cc}
\tabletypesize{\footnotesize}
\tablecolumns{7} 
\tablewidth{0pc}
\tablecaption{Best-fitting power laws to two-point correlation functions.} 
\tablehead{  
  & \multicolumn{2}{c}{Cep A R5} & \multicolumn{2}{c}{W75 N VLA 1} &\multicolumn{2}{c}{W75 N VLA 2}\\
\cline{2-3}\cline{4-5}\cline{6-7}
  & \colhead{Index $\alpha$~\tablenotemark{a}} & \colhead{Index $\alpha$~\tablenotemark{a}} & \colhead{Index $\alpha$~\tablenotemark{a}} & \colhead{Index $\alpha$~\tablenotemark{a}} & \colhead{Index $\alpha$~\tablenotemark{a}} &  \colhead{Index $\alpha$~\tablenotemark{a}}\\
  & small scales & large scales & small scales & large scales & small scales & large scales\\
 Epoch & (0.07--0.24~AU) & (0.5--70~AU) & (0.2--0.75~AU) & (1.3--1500~AU) & (0.2--0.6~AU) & (1.5--250~AU)
}
\startdata 
 1st & $-2.8\pm$0.6 & $-1.48\pm$0.15 & $-2.9\pm$0.9 & $-1.47\pm$0.16  & $-2.6\pm$0.4   & $-1.27\pm$0.13 \\
 2nd & $-2.4\pm$0.4 & $-1.56\pm$0.25 & $-4.3\pm$0.4 & $-1.49\pm$0.18  & $-2.44\pm$0.20 & $-1.37\pm$0.12\\
 3rd & $-3.0\pm$0.3 & $-1.41\pm$0.14 & $-3.2\pm$0.8 & $-1.45\pm$0.16  & $-2.8\pm$0.5   & $-1.28\pm$0.10\\
\enddata
\tablenotetext{a}{Two-point correlation functions for maser spots can be fitted by: $n_s(\Delta r)=n_0\Delta r^{\alpha}$}
\tablecomments{Uncertainties are 2$\sigma$.}
\end{deluxetable}

\begin{deluxetable}{ccc@{\extracolsep{1.5em}}cc@{\extracolsep{1.5em}}cc}
\tabletypesize{\footnotesize}
\tablecolumns{7} 
\tablewidth{0pc}
\tablecaption{Best-fitting power laws to velocity correlation functions.} 
\tablehead{  
  & \multicolumn{2}{c}{Cep A R5~\tablenotemark{a}} & \multicolumn{2}{c}{W75 N VLA 1~\tablenotemark{b}} &\multicolumn{2}{c}{W75 N VLA 2~\tablenotemark{c}}\\
\cline{2-3}\cline{4-5}\cline{6-7}
Epoch & Index $\alpha$~\tablenotemark{d}& Index $\alpha$~\tablenotemark{e} & Index $\alpha$~\tablenotemark{d} & Index $\alpha$~\tablenotemark{e} & Index $\alpha$~\tablenotemark{d} & Index $\alpha$~\tablenotemark{e} 
}
\startdata 
1st & $-0.04\pm$0.06 &      $-0.01\pm$0.06 & $0.13\pm$0.07 & $0.15\pm$0.08 & $0.14\pm$0.06 & $0.19\pm$0.05\\
2nd & $-0.02\pm$0.05 &  \phs $0.01\pm$0.06 & $0.14\pm$0.09 & $0.14\pm$0.09 & $0.12\pm$0.04 & $0.16\pm$0.04\\
3rd & $-0.03\pm$0.06 &  \phs $0.02\pm$0.08 & $0.31\pm$0.11 & $0.14\pm$0.08 & $0.15\pm$0.05 & $0.23\pm$0.04\\
\enddata
\tablenotetext{a}{Scale range: 0.07--70~AU.}
\tablenotetext{b}{Scale range: 0.2--1500~AU.}
\tablenotetext{c}{Scale range: 0.2--320~AU.}
\tablenotetext{d}{Index of the power-law fit to the rms velocity difference between maser spot pairs, \mbox{$V_s(\Delta r)\propto\Delta r^{\alpha}$}.}
\tablenotetext{e}{Index of the power-law fit to the median velocity difference between maser spot pairs, \mbox{$M_s(\Delta r)\propto\Delta r^{\alpha}$}.}
\tablecomments{Uncertainties are 2$\sigma$.}
\end{deluxetable}

\clearpage

\end{document}